\documentclass[prl,aps,footinbib,twocolumn,showpacs]{revtex4-1}
\usepackage{amssymb}
\usepackage{graphicx}
\usepackage{amsmath}
\usepackage{times}
\usepackage{color}
\usepackage{subfigure}
\usepackage{setspace}
\usepackage{bm}
\usepackage{braket}
\usepackage[bookmarks=false,colorlinks=true,linkcolor=blue,citecolor=blue]{hyperref}

\begin{document}

\title{ Optical spectrum analyzer with quantum limited noise floor}

\author
{ M. Bishof, X. Zhang, M. J. Martin, and Jun Ye }
\affiliation{JILA, National Institute of Standards and Technology and University of Colorado,
Department of Physics, University of Colorado, Boulder, Colorado 80309, USA.}

\date{\today}
\begin{abstract}
Interactions between atoms and lasers  provide the potential for unprecedented control of quantum states. Fulfilling this potential requires detailed knowledge of frequency noise in optical oscillators with state-of-the-art stability.
We demonstrate a technique that precisely measures the noise spectrum of an ultrastable laser using optical lattice-trapped $^{87}$Sr atoms as a quantum projection noise-limited reference.
We determine the laser noise spectrum from near DC to 100 Hz via the measured fluctuations in atomic excitation, guided by a simple and robust theory model.  The noise spectrum yields a 26(4) mHz linewidth at a central frequency of 429 THz, corresponding to an optical quality factor of $1.6\times10^{16}$.
This approach improves upon optical heterodyne beats between two similar laser systems by providing information unique to a single laser, and complements the traditionally used Allan deviation which evaluates laser performance at relatively long time scales.
We use this technique to verify the reduction of resonant noise in our ultrastable laser via feedback from an optical heterodyne beat.  Finally, we show that knowledge of our laser's spectrum allows us to accurately predict the laser-limited stability for optical atomic clocks.
\end{abstract}
\maketitle

The development of ultrastable frequency sources has paved the way for advances in fundamental tests of physics, primary frequency standards, precision spectroscopy, and quantum many-body systems.  However, the utility of a precision frequency source is limited by its instabilities. For this reason, many methods to rigourously characterize these instabilities have been developed \cite{rutman78}.   Ultrastable lasers pose a unique challenge to characterizing frequency instabilities because, until now, measurements of their performance required an optical herterodyne beat between two or more lasers \cite{young99,stoehr06,ludlow07,alnis08,dube09,jiang10}.
Single laser performance can be inferred from a three-cornered hat measurement \cite{zhao09,kessler12}, but valuable information about a laser's frequency noise power spectral density (PSD) \cite{cutler66} is limited in an optical beat by the less stable laser.

Optical lattice-trapped ${}^{87}$Sr atoms are uniquely suited for laser noise spectral analysis due to the ultranarrow linewidth and field insensitivity of the ${}^1S_0$ $(\Ket{g})$ to ${}^3P_0$ $(\Ket{e})$ clock transition as well as the low quantum projection noise (QPN) achievable with ensembles of many atoms.  To accomplish this, we adopt a technique similar to radio-frequency-based dynamical decoupling \cite{viola99} to manipulate the frequency noise sensitivity of this transition.  Previous implementations of dynamical decoupling manipulated radio-frequency transitions in quantum systems to eliminate \cite{biercuk09,kotler11} or analyze \cite{bylander11} environmental noise.  Here, the ${}^{87}$Sr clock transition is so insensitive to perturbations that we are able to measure the noise spectrum of the ultrastable laser used to excited it. To guide and interpret our experimental measurements, we develop a simple and robust theoretical framework that combines concepts from \cite{rutman78} with a model for atomic sensitivity to frequency fluctuations \cite{dick87,santarelli98,quessada03,note1}.  We compare experimentally measured fluctuations in atomic population to our theory and accurately determine the PSD of our laser.  As laser stability advances, we can continue to leverage the QPN-limited noise floor of this technique to analyze lasers with greater stability.

To model the frequency of our laser, we consider a fixed frequency with a small, time-dependent noise term:  $\omega_L(t)=\omega_{L_0}+\delta\omega(t)$. The instantaneous phase of the laser is given by $\phi_L(t) = \int_0^tdt'\omega_L(t') = \omega_{L_0}t+\int_0^tdt'\delta\omega(t') \equiv \omega_{L_0}t+\delta\phi(t)$.  The resulting Hamiltonian for a two level atom, with energy spacing $\hbar\omega_a$, driven by this laser is \cite{sup}
\begin{equation}
\frac{\hat{H}}{\hbar}=-\frac{\Omega(t)}{2}\left(
\begin{array}{cc}
0 & e^{i\delta\phi(t)} \\
e^{-i\delta\phi(t)} & 0
\end{array}
\right)-\frac{\overline{\Delta}}{2}\hat{\sigma}_z,
\label{eq:1}
\end{equation}
where $\overline{\Delta} \equiv \omega_{L_0} - \omega_a$, $\hat{\sigma}_z$ is a Pauli spin matrix, and $\Omega(t)$ is the Rabi frequency.  The chosen spectroscopy sequence determines the time dependence of $\Omega(t)$.  In the absence of other perturbations, the atom-light interaction can be engineered to filter laser noise.  For example, random fluctuations that occur on time scales that are fast compared to the atomic state evolution will average to zero.

\begin{figure}
    {\includegraphics[width =1\linewidth]{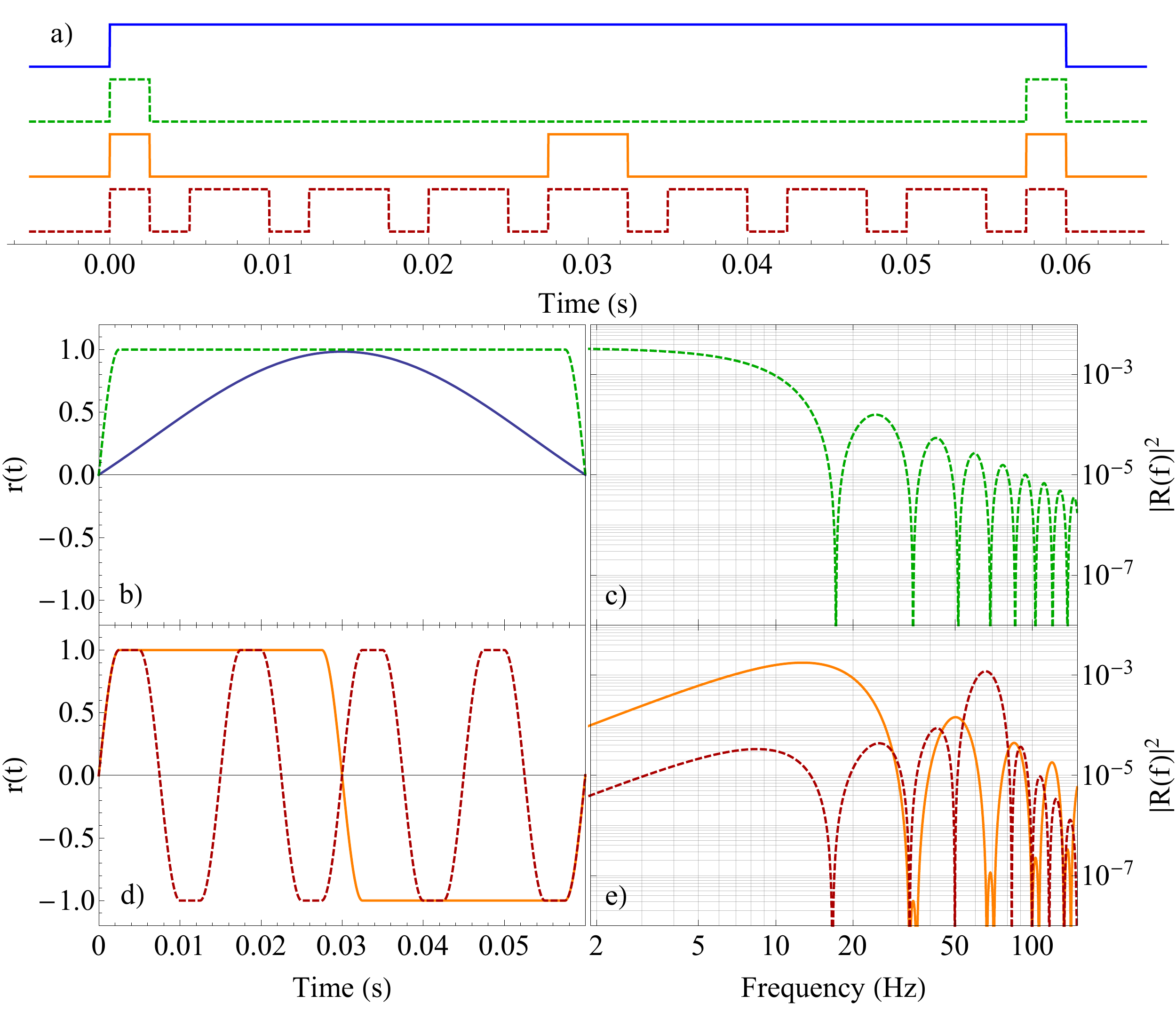}}
    \leavevmode
    \caption{ (color online)   Schematic diagrams,  $r(t)$, and $\left| R(f)\right|^2$ for different spectroscopy sequences.  An example total spectroscopy time of 60 ms is shown.  The schematic diagrams shown in (a) signify when $\Omega(t)$ is nonzero by their ``high'' value.  For Rabi spectroscopy, the nonzero $\Omega$ is selected so the total pulse area is $\pi$ ($\pi$ pulse).  For the other spectroscopy sequences, the nonzero value of $\Omega$ is $\pi/(0.005)$ rad/s.  The diagrams are offset in height for clarity and are ordered top to bottom:  Rabi, Ramsey, one-echo, and seven-echo..  Plots of $\left| R(f)\right|^2$ in (c) and (e) are calculated from the corresponding $r(t)$ curves plotted in (b) and (d).  In (a), (b) and (c) solid blue (dashed green) lines correspond to Rabi (Ramsey) spectroscopy sequences.  In (a), (d) and (e) solid orange (dashed red) lines correspond to one-echo (seven-echo) spectroscopy sequences.}   \label{fig1}
\end{figure}

To measure the effect of laser frequency fluctuation on the atoms, we observe fluctuations in the population imbalance between $\Ket{g}$ and $\Ket{e}$.  For a general state $\Ket{\psi}=a\Ket{g}+b\Ket{e}$, the population imbalance is defined as $\mathbb{P} \equiv bb^\ast - aa^\ast$.  We can express $\mathbb{P}$ in terms of the time-dependent laser detuning as
\begin{equation}
    \mathbb{P}(\tau) = \mathbb{P}_0 + \int_{0}^\tau dt\, r(t)\,\Delta(t),
    \label{eq:6}
\end{equation}
where $\mathbb{P}_0$ is the initial imbalance, $\tau$ is the total spectroscopy time, $\Delta(t)\equiv\overline{\Delta}+\delta\omega(t)$, and $r(t)$ is the impulse response \cite{rutman78}, commonly referred to as the sensitivity function \cite{santarelli98,quessada03}.  The sensitivity function, and its Fourier transform $R(f)$, are determined by the chosen spectroscopy sequence.  As we apply different spectroscopy sequences, fluctuations in $\mathbb{P}$ correspondingly reveal laser instabilities at different Fourier frequencies, as illustrated by the shifting spectral response of $\left|R(f)\right|^2$ in Fig.\ \hyperref[fig1]{\ref{fig1}(c) and 1(e)}.

Equation (\ref{eq:6}) rigorously connects $\mathbb{P}$ to $\Delta(t)$ and, to quantify fluctuations in $\mathbb{P}$, we consider its variance, $I^2\equiv \left\langle\mathbb{P}^2\right\rangle - \left\langle\mathbb{P}\right\rangle^2$, which can be expressed as \cite{sup}
\begin{equation}
   I^2 = (2\pi)^2\int_{0}^\infty df\, S_\nu(f)\,\left|R(f)\right|^2.
    \label{eq:12}
\end{equation}
Here, $S_\nu(f)$ is the single sided frequency noise PSD of the laser in units of Hz${}^2$/Hz.  Figure \ref{fig1} shows a schematic diagram for the spectroscopy sequences we use along with their corresponding sensitivity functions, $r(t)$ and $\left|R(f)\right|^2$.

For Rabi and Ramsey spectroscopy, the calculated value for $I$ diverges since thermal noise, a fundamental limit to $S_\nu(f)$ at low frequency, has an $f^{-1}$ character.  For these measurements, we use the Allan deviation \cite{allan66} to characterize fluctuations in $\mathbb{P}$.  In particular, we consider the two-sample Allan variance, defined as
\begin{equation}
   I_{(2)}^2 \equiv \frac{1}{2}\left\langle(\mathbb{P}_{i+1}-\mathbb{P}_i)^2\right\rangle,
   \label{eq:13}
\end{equation}
where the index $i$ signifies the $i$th measurement of $\mathbb{P}$.  In the treatment of multiple measurements we consider the sensitivity function as periodic with a period equal to the experimental cycle time $T_c$.  For this work, $T_c$ is approximately $1+\tau$ s.  $I_{(2)}^2$ can also be expressed in terms of $S_\nu(f)$ as follows \cite{sup}:
\begin{equation}
   I_{(2)}^2 = (2\pi)^2\int_{0}^\infty df\, S_\nu(f)\,2\sin^2(\pi f T_c)\left|R(f)\right|^2.
    \label{eq:15}
\end{equation}
Although the calculated value of $I_{(2)}$ remains finite for all experimental conditions, it does not properly account for coherent vibrational or electronic noise that exists on our laser at frequencies above 20 Hz.  This noise is aliased onto our measurements and leads to regular, slow oscillations of the measured $\mathbb{P}$. To capture the effect of coherent noise, we use $I$ to characterize echo pulse sequences.  Rabi and Ramsey sequences do not suffer from this aliasing because they do not have significant sensitivity to noise above 20 Hz for the spectroscopy times we use.

\begin{figure*}
{\includegraphics[width =1\linewidth]{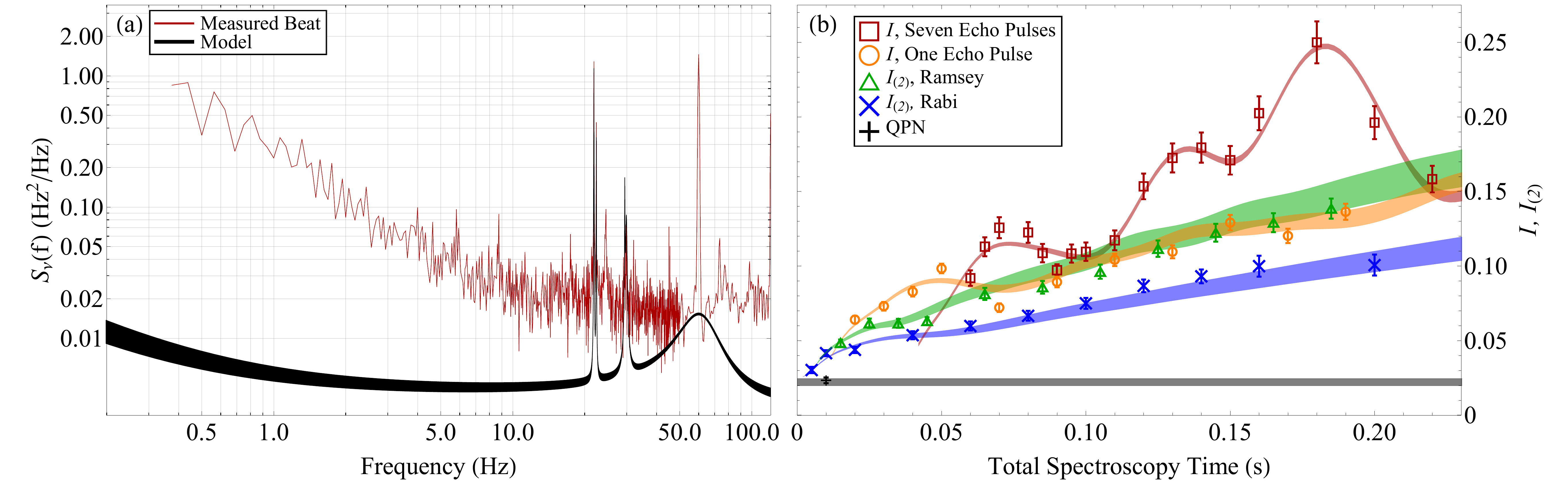}}
 \leavevmode
 \caption{ (color online)   (a) $S_\nu(f)$ measured from a beat between the $\alpha$  and  $\beta$ lasers is plotted in red.  A model $S_\nu(f)$ for the $\alpha$ laser is plotted as a black band (see the text).
 (b) Measured values of $I$ or $I_{(2)}$ are plotted as a function of total spectroscopy time for the spectroscopy sequences shown in Fig.\ \ref{fig1}.  Colored bands represent calculated values of $I$ or $I_{(2)}$ using the model $S_\nu(f)$ in (a) with QPN added in quadrature.  A gray band indicates the mean calculated QPN for all experimental data.  The black cross represents a measurement of the QPN (see the text).}   \label{fig2}
\end{figure*}

Our experimental setup follows that of our Sr clock \cite{bishof11,campbell08}.  Between 2000 and 3000 ${}^{87}$Sr atoms are cooled to about 2 $\mathrm{\mu}$K in a one-dimensional optical lattice and nuclear spin polarized into the ground ${}^1S_0$ $m_F=9/2$ state. The optical lattice is kept near the magic wavelength \cite{ye08} for the $\Ket{g}$ to $\Ket{e}$ clock transition.  Lattice-trapped atoms are excited with 698 nm light according to the spectroscopy sequences shown in Fig.\ \hyperref[fig1]{\ref{fig1}(a)}.  The clock light propagates along the strongly confined axis of the lattice so that it probes the atoms in the well-resolved sideband regime \cite{wineland98,blatt09}, free from Doppler and recoil effects.  Finally, the numbers of atoms in $\Ket{g}$ and $\Ket{e}$ are measured to determine $\mathbb{P}$.

Using this setup, we have resolved 0.5 Hz spectral features \cite{swallows12,martin13} and demonstrated the most stable optical clock \cite{nicholson12}.  These results are enabled by the ultrastable laser that addresses the clock transition (hereafter termed ``$\alpha$ laser'').  The stability of the $\alpha$ laser is at its thermal noise limit of $1\times10^{-16}$ fractional frequency units for $\sim$1 to 1000 s.
We can look for noise features at higher frequencies using an optical beat with a second laser (hereafter termed ``$\beta$ laser'').  The $\beta$ laser has demonstrated thermal noise-limited stability at the $10^{-15}$ fractional frequency level \cite{ludlow07}.  The PSD of the optical beat (Fig.\ \hyperref[fig2]{\ref{fig2}(a)}) is limited by thermal noise in the $\beta$ laser out to Fourier frequencies of 10 Hz, beyond which it becomes limited by the noise floor of the detector; however, discrete features exist above this floor.  Pairs of narrow noise peaks are visible near 22 and 30 Hz.  Additionally, noise peaks are consistently measured at 24 and 60 Hz.  The 60 Hz peak is dominated by detector noise and the 24 Hz peak is visible in previous beat measurements between two $\beta$ lasers \cite{ludlow07}.

Spectroscopy sequences are designed such that the measured $\mathbb{P}$ is sensitive to laser noise.  For Rabi spectroscopy, we detune the $\alpha$ laser from resonance by the half width at half maximum (HWHM) of the Rabi line shape and apply a $\pi$ pulse.  For Ramsey spectroscopy, we tune the laser exactly on resonance and apply two $\pi/2$ pulses separated in time.  We shift the phase of the final $\pi/2$ pulse by $\pi/2$ radians relative to the initial $\pi/2$ pulse, which is equivalent to detuning by the HWHM of the central Ramsey fringe in the absence of phase shifts.  The echo pulse sequences add to the Ramsey sequence a number of $\pi$ pulses such that the free evolution times between pulses are equal.  We switch the phase of the laser by $\pi$ rad between adjacent echo pulses so that pulse area errors cancel.  The echo pulse sequences act as a bandpass filter peaked at $(n+1)/(2\tau)$ Hz, where $n$ is the number of echo pulses. One can intuitively understand this behavior from the sensitivity functions in Fig.\ \hyperref[fig1]{\ref*{fig1}(d)}, which are periodic at this frequency.  Figure \hyperref[fig1]{\ref*{fig1}(e)} explicitly demonstrates this frequency sensitivity.

We use 80 consecutive measurements of $\mathbb{P}$ to estimate the raw standard(pair) deviation and its statistical uncertainty then divide by the measured contrast to get $I$($I_{(2)}$).  The contrast is determined by a fit to the measured excitation versus detuning for Rabi spectroscopy or a fit to measured oscillations in excitation as the phase of the final pulse is scanned for other sequences.   Figure \hyperref[fig2]{\ref{fig2}(b)} shows measured values of $I$ or $I_{(2)}$ for different spectroscopy sequences as a function of total spectroscopy time.  Each data point represents a weighted mean of at least four measurements and error bars are estimated from the variance of the weighted mean.  Spectroscopy times are investigated in a random order to avoid systematic drifts.  Each data point consists of measurements separated by several hours to ensure consistency of the data.

The spectroscopy sequences we use are chosen to measure different Fourier components of $S_\nu(f)$ and demonstrate the utility of dynamical decoupling.  The peak frequency sensitivity of the one-echo pulse data ranges between 5 and 50 Hz; however, individual noise components cannot be identified.  By increasing the number of echo pulses to seven, we clearly resolve three peaks in the measured values of $I$, centered at 0.070, 0.135, and 0.180 s of total spectroscopy time.  The peak centered at 0.135 s originates from alternating current motors in our lab operating near 30 Hz.  The peak at 0.180 s corresponds to an acoustic resonance of the lab at 22 Hz.  The width of the peak at 0.070 s corresponds to a frequency width that is broader than the resolution of the seven-echo pulse sequence (roughly $\tau^{-1}$).  It contains multiple unresolved noise components corresponding to electrical noise at 60 Hz and acoustic noise near 40 and 80 Hz, which was previously observable in the optical heterodyne beat prior to the installation of an acoustic isolation box around the $\alpha$ laser.  Lasers are also subject to white noise (no dependence on Fourier frequency) and noise proportional to $1/f$ originating from electronic and thermal noise respectively.  At the magnitudes we extract from the experimental data, these noise components have a negligible effect on the calculated values of $I$ for seven-echo pulse sequences.  In contrast, calculated values of $I_{(2)}$ for Rabi and Ramsey pulse sequences depend primarily on the magnitudes of white and $1/f$ noise since their $\left|R(f)\right|^2$ decreases with increasing $f$. The agreement between these two sequences is used to bound the uncertainty in the magnitudes of white and $1/f$ noise.

To determine $S_\nu(f)$ for the $\alpha$ laser we fit measured values of $I$ and $I_{(2)}$ to theoretical calculations using a single model $S_\nu(f)$ in Eqs. (\ref{eq:12}) and (\ref{eq:15}).  The functional form of the model is
\begin{equation}
S_\nu(f) = h_\mathrm{white}+\frac{h_\mathrm{thermal}}{f}+\sum_{i=1}^N\frac{h_i}{1+\left(\frac{f-f_i}{\Gamma_i/2}\right)^2},
\label{eq:16}
\end{equation}
where $h_i$, $f_i$, and $\Gamma_i$ are the magnitude, frequency, and full width at half maximum (FWHM) for the $i^{th}$ noise resonance.  All $h_i$ are fit to the seven-echo pulse data and the Rabi and Ramsey data are used to simultaneously fit $h_{\mathrm{white}}$ and $h_{\mathrm{thermal}}$.  We determine $h_\mathrm{thermal}=1.5(4)\times10^{-3}$ Hz${}^2$, consistent with predicted thermal noise \cite{swallows12}, and $h_\mathrm{white}=3.3(3)\times10^{-3}$ Hz${}^2/$Hz.  Widths and frequencies of discrete noise peaks are chosen to be consistent with the optical beat.  The parameters for these resonances are given in \cite{sup}. We note that widths and frequencies could be identified without the aid of the optical beat, as demonstrated by the distinct peaks in Fig.\ \hyperref[fig2]{\ref{fig2}(b)}, but the resolution would be limited to $\sim1/\tau$.  Although an exact relationship between $S_\nu(f)$ and a FWHM linewidth exists \cite{elliott82}, an analytic expression for this relationship does not exist with $1/f$ frequency noise.  Here, the observed linewidth depends on the measurement time \cite{didomenico10}.  By accounting for a finite measurement time, we numerically calculate the minimum observable $\alpha$ laser linewidth to be 26(4) mHz \cite{sup}.

For each data point, the QPN is calculated for the measured number of atoms and the mean excitation fraction. The mean QPN for each sequence is added in quadrature with the calculated $I$ and $I_{(2)}$ to more accurately represent experimental data.  These quantities are plotted as colored bands in Fig.\ \hyperref[fig2]{\ref{fig2}(b)} where the extent of the band corresponds to the uncertainty of the model $S_\nu(f)$.
The mean and standard deviations of all calculated QPN values are represented in Fig.\ \hyperref[fig2]{\ref{fig2}(b)} as a gray band.  QPN is experimentally measured by the standard deviation of $\mathbb{P}$ following a 5 ms, resonant, $\pi/2$ pulse. The measured and calculated QPN are consistent.

 \begin{figure}
{\includegraphics[width =1\linewidth]{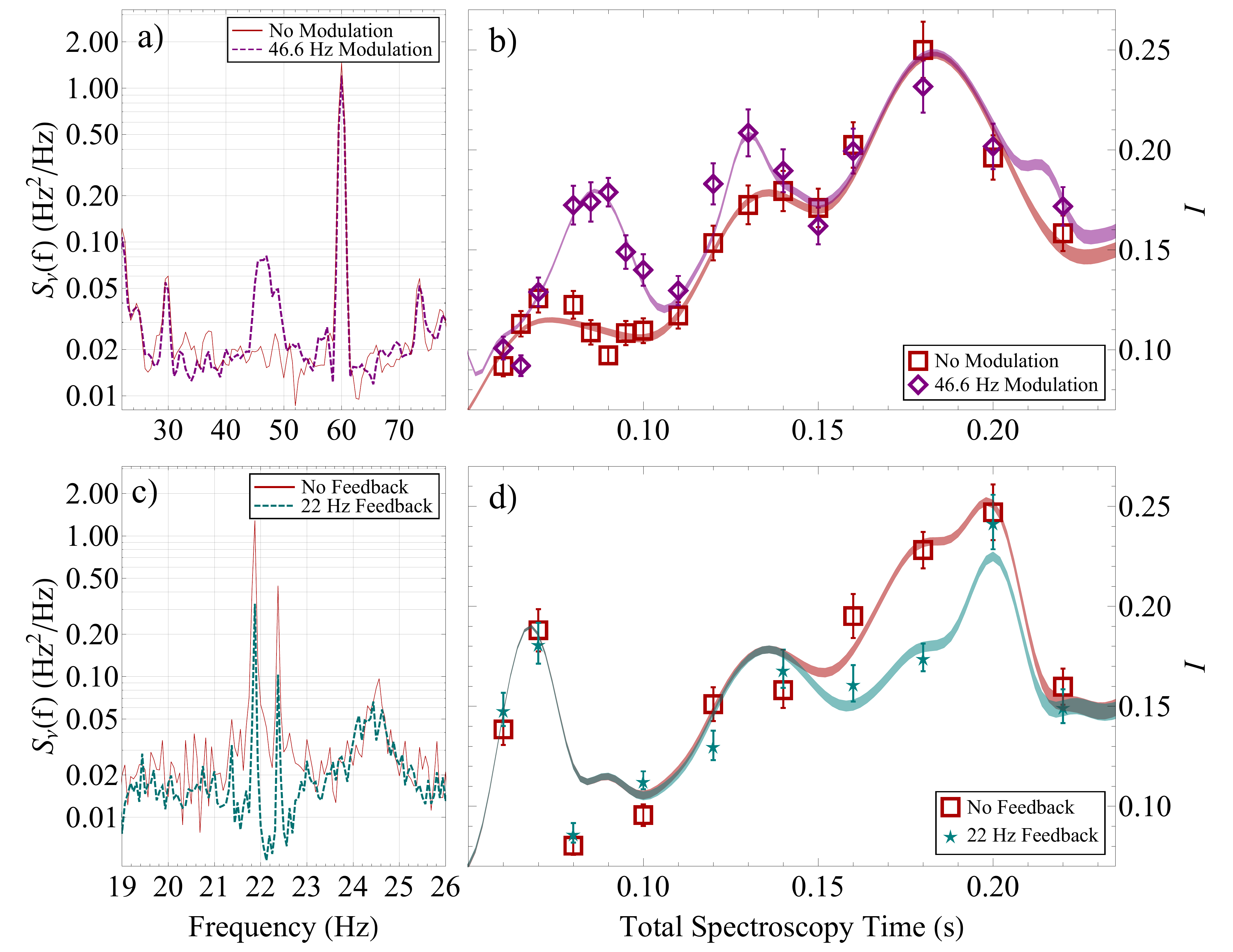}}
 \leavevmode
 \caption{ (color online)   (a) Measured $S_\nu(f)$ of an optical beat between the $\alpha$  and $\beta$ lasers and (b) measured and calculated  $I$ for seven-echo pulse sequences with and without modulation at 46.6 Hz.  (c) Measured $S_\nu(f)$ of an optical beat between the $\alpha$  and $\beta$ lasers and (d) measured and calculated  $I$ for seven-echo pulse sequences with and without feedback cancelation of 22 Hz noise.  (d) Also demonstrated is the appearance of 60 Hz noise due to a malfunctioning signal generator that increases the measured $I$ at 0.06 and 0.2 s of total spectroscopy time.}   \label{fig3}
\end{figure}

To further test our theory we intentionally add noise to the $\alpha$ laser.  White noise is passed through a bandpass filter at 46.6 Hz with 2 Hz bandwidth and used to frequency modulate the $\alpha$ laser with an acoustic optical modulator (AOM). Figure \hyperref[fig3]{\ref{fig3}(a)} demonstrates the effect of the modulation on the optical beat between the $\alpha$  and $\beta$ lasers.  By adding noise into our model $S_\nu(f)$, corresponding to the 1st and 2nd order contributions of the modulation, we can fully account for measured values of $I$ with modulation.  Figure \hyperref[fig3]{\ref{fig3}(b)} shows calculated and measured values of $I$ for seven-echo pulse spectroscopy with and without 46.6 Hz modulation.

In addition to using the optical beat to validate atomic measurements, we also harness the information within the beat to reduce discrete noise features in the $\alpha$ laser.  We filter the beat with a bandpass at 22 Hz having subhertz bandwidth. This signal is inverted and fed back onto the $\alpha$ laser with an AOM.
We can observe the effect of feedback on the beat (Fig.\ \hyperref[fig3]{\ref{fig3}(c)}), although we truly demonstrate the effectiveness of this technique by observing a reduction in $I$ for $\tau$ between 0.15 and 0.20 s when feedback is active (Fig.\ \hyperref[fig3]{\ref{fig3}(d)}).  To reproduce the measured $I$ with feedback, the magnitude of 22 Hz noise needed to be reduced by 60\% in the model $S_\nu(f)$ compared to the condition without feedback modulation.

 \begin{figure}
{\includegraphics[width =1\linewidth]{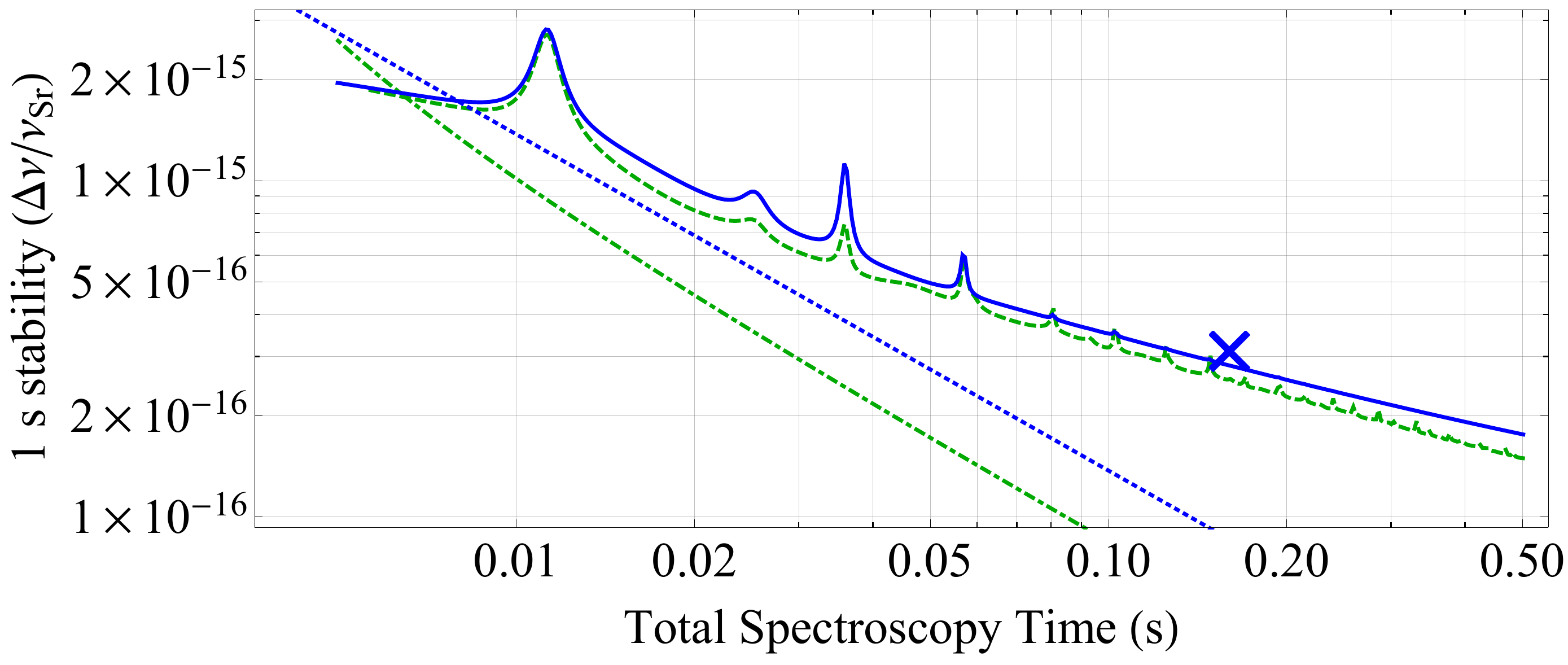}}
 \leavevmode
 \caption{ (Color online)   The solid blue (dashed green) line represents the 1 s stability limit for an optical clock due to the Dick effect using Rabi (Ramsey) spectroscopy.  The blue dotted (green dot-dashed) line represents the QPN stability limit at 1 s for Rabi (Ramsey) spectroscopy assuming a collection of 2000 uncorrelated atoms. Ramsey spectroscopy assumes 2.5 ms $\pi/2$ pulses.  The blue ``x'' denotes the single clock 1 s stability from Ref. \cite{nicholson12}.}   \label{fig4}
\end{figure}

Having developed an accurate model for the $\alpha$ laser's $S_\nu(f)$, we can predict the stability this laser can achieve when used in an optical atomic clock.  Here, the laser's frequency is slaved to the clock transition by periodic interrogation. The stability is limited by the Dick effect \cite{dick87,santarelli98,quessada03}, whereby periodic interrogation creates sensitivity to laser noise at harmonics of $1/T_c$. Figure \ref{fig4}, plots the one second stability limit due to the Dick effect for Rabi and Ramsey spectroscopy as a function of $\tau$, using the lower limit of the model $S_\nu(f)$.  We assume a typical $T_c$ of $857.5+\tau$ ms.  We find that for Rabi spectroscopy with $\tau=160$ ms, the Dick effect limits clock stability to $2.8\times10^{-16}/\sqrt{\tau}$ in fractional frequency units. For a comparison of two uncorrelated clocks, one operating with 1000 atoms and one operating with 2000 atoms, we predict a stability of $4.6\times10^{-16}/\sqrt{\tau}$ which is within $5\%$ of the achieved stability in Ref. \cite{nicholson12}.

We thank R.\ Ozeri for stimulating discussions on dynamical decoupling and J.\ K.\ Thompson and Z.\ Chen for parallel work and discussions on the treatment of oscillator phase noise in the Bloch vector picture \cite{chen12}.  We thank T.\ L.\ Nicholson, B.\ J.\ Bloom, J.\ R.\ Williams, W.\ Zhang and S.\ L.\ Campbell for useful discussions.  We acknowledge funding support for this work by DARPA QuASAR, NIST, and NSF.\,  M.\ B.\ acknowledges support from NDSEG.


\begin{thebibliography}{22}
\bibitem{note1}
We note that our treatment is consistent with that in \cite{chen12}.
\end{thebibliography}
\end{document}


\title{Supplementary online material to the manuscript:\\``Optical spectrum analyzer with  quantum limited noise floor'' }

\author
{ M. Bishof, X. Zhang, M. J. Martin, and Jun Ye }
\affiliation{JILA, National Institute of Standards and Technology and University of Colorado,
Department of Physics, University of Colorado, Boulder, Colorado 80309, USA.}
\maketitle

\section{Derivation of the detuning-dressed basis Hamiltonian}

We start with the Schr\"{o}dinger equation for a two-level atom driven by a laser field with a frequency that can vary in time.  This is most simply treated using the instantaneous phase of the laser, $\phi_L(t)=\int_0^t\omega_L(t)dt$, where $\omega_L(t)$ is the instantaneous frequency of the laser.  For a general state $\ket{\psi}=c_g(t)\ket{g}+c_e(t)\ket{e}$, we get the coupled differential equations:
\begin{equation}
\begin{array}{rcl}
i\dot{c}_e(t)&=&\frac{\omega_a}{2}c_e(t)-\frac{\Omega(t)}{2}\left(e^{i\phi_L(t)}+e^{-i\phi_L(t)}\right)c_g(t), \\
\\
i\dot{c}_g(t)&=&-\frac{\omega_a}{2}c_g(t)-\frac{\Omega(t)}{2}\left(e^{i\phi_L(t)}+e^{-i\phi_L(t)}\right)c_e(t).
\end{array}
\label{eq:1a}
\end{equation}
Here, $\Omega(t)$ is the Rabi frequency and $\omega_a$ is the frequency of the $\ket{g}$ to $\ket{e}$ transition.

To simplify the solutions to these equations, we substitute the complex state coefficients with
\begin{equation}
\tilde{c}_e(t)\equiv e^{i\omega_at/2}c_e(t) \quad\mathrm{and} \quad \tilde{c}_g(t)\equiv e^{-i\omega_at/2}c_g(t).
\label{eq:2a}
\end{equation}
Using this substitution, (\ref{eq:1a}) becomes
\begin{equation}
\begin{array}{lr}
\frac{\omega_a}{2}e^{-i\omega_at/2}\tilde{c}_e(t)+ie^{-i\omega_at/2}\dot{\tilde{c}}_e(t)& \\
=\frac{\omega_a}{2}e^{-i\omega_at/2}\tilde{c}_e(t)-\frac{\Omega(t)}{2}\left(e^{i\phi(t)}+e^{-i\phi(t)}\right)e^{i\omega_at/2}\tilde{c}_g(t), \\
\\
-\frac{\omega_a}{2}e^{i\omega_at/2}\tilde{c}_g(t)+ie^{i\omega_at/2}\dot{\tilde{c}}_g(t)& \\
=-\frac{\omega_a}{2}e^{i\omega_at/2}\tilde{c}_g(t)-\frac{\Omega(t)}{2}\left(e^{i\phi(t)}+e^{-i\phi(t)}\right)e^{-i\omega_at/2}\tilde{c}_e(t).
\end{array}
\label{eq:3a}
\end{equation}
We can cancel an overall phase factor of $e^{-(+)i\omega_at/2}$ from the top(bottom) equation in (\ref{eq:3a}) and simplify to get
\begin{equation}
\begin{array}{rcl}
i\dot{\tilde{c}}_e(t)&=&
-\frac{\Omega(t)}{2}\left(e^{i(\phi_L(t)+\omega_at)}+e^{-i(\phi_L(t)-\omega_at)}\right)\tilde{c}_g(t), \\
\\
i\dot{\tilde{c}}_g(t)&=&
-\frac{\Omega(t)}{2}\left(e^{i(\phi_L(t)-\omega_at)}+e^{-(i\phi_L(t)+\omega_at)}\right)\tilde{c}_e(t).
\end{array}
\label{eq:4a}
\end{equation}

In order to proceed, we assume that the instantaneous laser frequency is the sum of a constant frequency and a small noise term that varies in time, $\omega_L(t) = \omega_{L_0} + \delta\omega(t)$.  Then we can write the instantaneous phase of the laser as $\phi_L(t) = \omega_{L_0}t+\int_0^t\delta\omega(t')dt'\equiv\omega_{L_0}t+\delta\phi(t)$.  Substituting this into (\ref{eq:4a}), we get:
\begin{equation}
\begin{array}{ll}
i\dot{\tilde{c}}_e(t)&=-\frac{\Omega(t)}{2}\times \\
&\left(e^{i(\omega_{L_0}t+\delta\phi(t)+\omega_at)}+e^{-i(\omega_{L_0}t+\delta\phi(t)-\omega_at)}\right)\tilde{c}_g(t), \\
\\
i\dot{\tilde{c}}_g(t)&=-\frac{\Omega(t)}{2}\times \\
&\left(e^{i(\omega_{L_0}t+\delta\phi(t)-\omega_at)}+e^{-i(\omega_{L_0}t+\delta\phi(t)+\omega_at)}\right)\tilde{c}_e(t).
\end{array}
\label{eq:5a}
\end{equation}
Since, $\delta\phi(t)$ is assumed to be small, we can make the usual rotating wave approximation where we neglect terms that rotate at the sum of $\omega_{L_0}$ and $\omega_a$.  The resulting equations are
\begin{equation}
\begin{array}{ll}
i\dot{\tilde{c}}_e(t)&=-\frac{\Omega(t)}{2}e^{-i(\overline{\Delta}t+\delta\phi(t))}\tilde{c}_g(t), \\
\\
i\dot{\tilde{c}}_g(t)&=-\frac{\Omega(t)}{2}e^{i(\overline{\Delta}t+\delta\phi(t))}\tilde{c}_e(t).
\end{array}
\label{eq:6a}
\end{equation}
where, $\overline{\Delta}\equiv\omega_{L_0}-\omega_a$.

Another basis change is necessary before arriving at the final result, so we express (\ref{eq:6a}) in terms of new state coefficients defined as
\begin{equation}
b_e(t)\equiv e^{-i\overline{\Delta} t/2}\tilde{c}_e(t) \quad\mathrm{and} \quad b_g(t)\equiv e^{i\overline{\Delta} t/2}\tilde{c}_g(t).
\label{eq:7a}
\end{equation}

After canceling overall phase factors we get the following differential equations.
\begin{equation}
\begin{array}{ll}
i\dot{b}_e(t)&=-\frac{\Omega(t)}{2}e^{-i\delta\phi(t)}b_g(t)+\frac{\overline{\Delta}}{2}b_e(t), \\
\\
i\dot{b}_g(t)&=-\frac{\Omega(t)}{2}e^{i\delta\phi(t)}b_e(t)-\frac{\overline{\Delta}}{2}b_g(t).
\end{array}
\label{eq:9a}
\end{equation}

This gives that the Hamiltonian in this basis is
\begin{equation}
\frac{\hat{H}}{\hbar}=-\frac{\Omega(t)}{2}\left(
\begin{array}{cc}
0 & e^{i\delta\phi(t)} \\
e^{-i\delta\phi(t)} & 0
\end{array}
\right)-\frac{\overline{\Delta}}{2}\hat{\sigma}_z,
\label{eq:10a}
\end{equation}
where $\hat{\sigma}_z$ is a pauli spin matrix.

\section{Calculation of the sensitivity function}

Before we can calculate the variance of $\mathbb{P}$ using Eqns.\ (3) or (5) from the main text, we will first need to calculate the sensitivity function, r(t), for a given spectroscopy sequence.  This quantity describes the atomic response to frequency deviations of the form $\delta\omega(t,t_0) = \Delta\phi\,\delta(t-t_0)$, where $\delta(t)$ is the Dirac delta function and $\Delta\phi$ is infinitesimally small.  Using the definition of $\delta\phi(t)$ from the main text, we see that frequency fluctuations of this form are equivalent to instantaneous steps in $\delta\phi(t)$ at time $t_0$. Adding finite steps in $\delta\phi(t)$ can also account for deliberate phase shifts of the laser which we implement during spectroscopy.

Inserting the above form of $\delta\omega(t,t_0)$ into Eqn.\ (2) from the main text and differentiating with respect to $\Delta\phi$, we get
\begin{equation}
    \left.\frac{\partial\mathbb{P}}{\partial\Delta\phi}\right\rvert_{\Delta\phi=0} = \int_{0}^\tau dt\, r(t)\, \delta(t-t_0) = r(t_0).
\label{eq:7}
\end{equation}
We calculate $\mathbb{P}$ using Eqn.\ (1) from the main text for a particular spectroscopy sequence, $\Omega(t)$, and with an added phase step of $\Delta\phi$ occurring at time $t_0$.  Finally, $r(t)$ is calculated by allowing $t_0$ to vary throughout the spectroscopy sequence.

\section{Derivation of variance}

Here we derive the variance of $\mathbb{P}$, starting with Eqn. (2) in the main text.  We assume that all measurements are made near $\mathbb{P} = 0$ so it will be sufficient to calculate the quantity
\begin{equation}
    \left\langle\mathbb{P}^2\right\rangle =\left\langle\int_{-\infty}^\infty\int_{-\infty}^\infty dt_1\,dt_2\,\Delta(t_1)\,r(t_1)\,\Delta(t_2)\,r(t_2)\right\rangle.
    \label{eq:1}
\end{equation}
This expression can be simplified using the definition of the convolution operation along with the definition of the autocorrelation function,
\begin{equation}
    \mathcal{R}_f(t_1-t_2) = \left\langle f(t_1)f^\ast(t_2)\right\rangle,
    \label{eq:2}
\end{equation}
where $R_f$ is the autocorrelation function of $f$.  Then, equation (\ref{eq:1}) simplifies to
\begin{equation}
     \left\langle\mathbb{P}^2\right\rangle = \int_{-\infty}^\infty dt\,(\mathcal{R}_\Delta\,\star \,r)(t)\,r(t),
     \label{eq:3}
\end{equation}
where ``$\star$'' represents convolution.  Since an oscillator's noise properties are typically characterized in frequency space via a power spectral density, we employ Parseval's theorem, which relates the integral of two complex valued functions $x(t)$ and $y(t)$ of time to the integral of their Fourier transforms $X(f)$ and $Y(f)$ as follows
\begin{equation}
    \int_{-\infty}^\infty dt\, x(t)\,y^\ast(t) = \int_{-\infty}^\infty df\, X(f)\,Y^\ast(f).
    \label{eq:4}
\end{equation}
Here, $f^\ast$ denotes the complex conjugate of $f$.
Since r(t) is real valued, we get that
\begin{equation}
      \int_{-\infty}^\infty dt\,(\mathcal{R}_\Delta\,\star \,r)(t)\,r(t)=\int_{-\infty}^\infty df\,\mathcal{F}\left[(\mathcal{R}_\Delta\,\star \,r)\right](f)\,R^\ast(f),
     \label{eq:5}
\end{equation}
where $\mathcal{F}$ is the fourier transform operation and $R(f)$ is the Fourier transform of $r(t)$.  The convolution theorem states that the Fourier transform of a convolution of two functions is equal to the product of the Fourier transforms of the two functions.  Additionally, the Wiener-Khinchin theorem states that the power spectral density (PSD) of a function, $f$,  is equal to the Fourier transform of the autocorrelation function of $f$.  Thus, we get that
\begin{equation}
    \left\langle\mathbb{P}^2\right\rangle = \int_{-\infty}^\infty df\, S_\Delta(f)\,\left|R(f)\right|^2,
    \label{eq:6}
\end{equation}
where $S_\Delta(f)$ is the double sided PSD of the laser in units of (rad/s)${}^2/$Hz.  We can express the variance in terms of the more commonly used single sided PSD with units of Hz${}^2/$Hz to obtain the final result
\begin{equation}
   I^2 = \left\langle\mathbb{P}^2\right\rangle = (2\pi)^2\int_{0}^\infty df\, S_\nu(f)\,\left|R(f)\right|^2.
    \label{eq:7}
\end{equation}
Here, we have use the relationship, $S_\Delta(f)=(2\pi)^2 S_\nu(f)/2$ for $f\geq0$.

\section{Derivation of the two sample Allan variance}
We begin with the definition for the two-sample Allan variance:
\begin{equation}
   I_{(2)}^2 \equiv \frac{1}{2}\left\langle(\mathbb{P}_{i+1}-\mathbb{P}_i)^2\right\rangle,
   \label{eq:8}
\end{equation}
Then the expression for the two-sample Allan variance is completely analogous to (\ref{eq:1}) under the substitution
\begin{equation}
   r(t) \rightarrow r_{(2)}(t) \equiv r(t) - r(t-T_c).
   \label{eq:9}
\end{equation}
Since the Fourier transform operation is linear and a shift in time of $T_c$ simply adds a phase factor of $e^{-2\pi\,i\,T_c\,f}$ to the Fourier transform of $r(t)$, we get that
\begin{equation}
\begin{array}{l}
   \mathcal{F}\left[r_{(2)}(t)\right] = \mathcal{F}\left[r(t)\right] - \mathcal{F}\left[r(t-T_c)\right] \\
   \\
   =2ie^{-\pi\,i\,T_c\,f}\,\sin{\left(\pi\,T_c\,f\right)}R(f).
   \end{array}
   \label{eq:10}
\end{equation}
This shows that $I_{(2)}^2$ can be expressed in terms of $R(f)$ as
\begin{equation}
   I_{(2)}^2 = (2\pi)^2\int_{0}^\infty df\, S_\nu(f)\,2\sin^2(\pi f T_c)\left|R(f)\right|^2.
    \label{eq:11}
\end{equation}

\section{Calculation of the laser linewidth}

 \begin{figure}
{\includegraphics[width =1\linewidth]{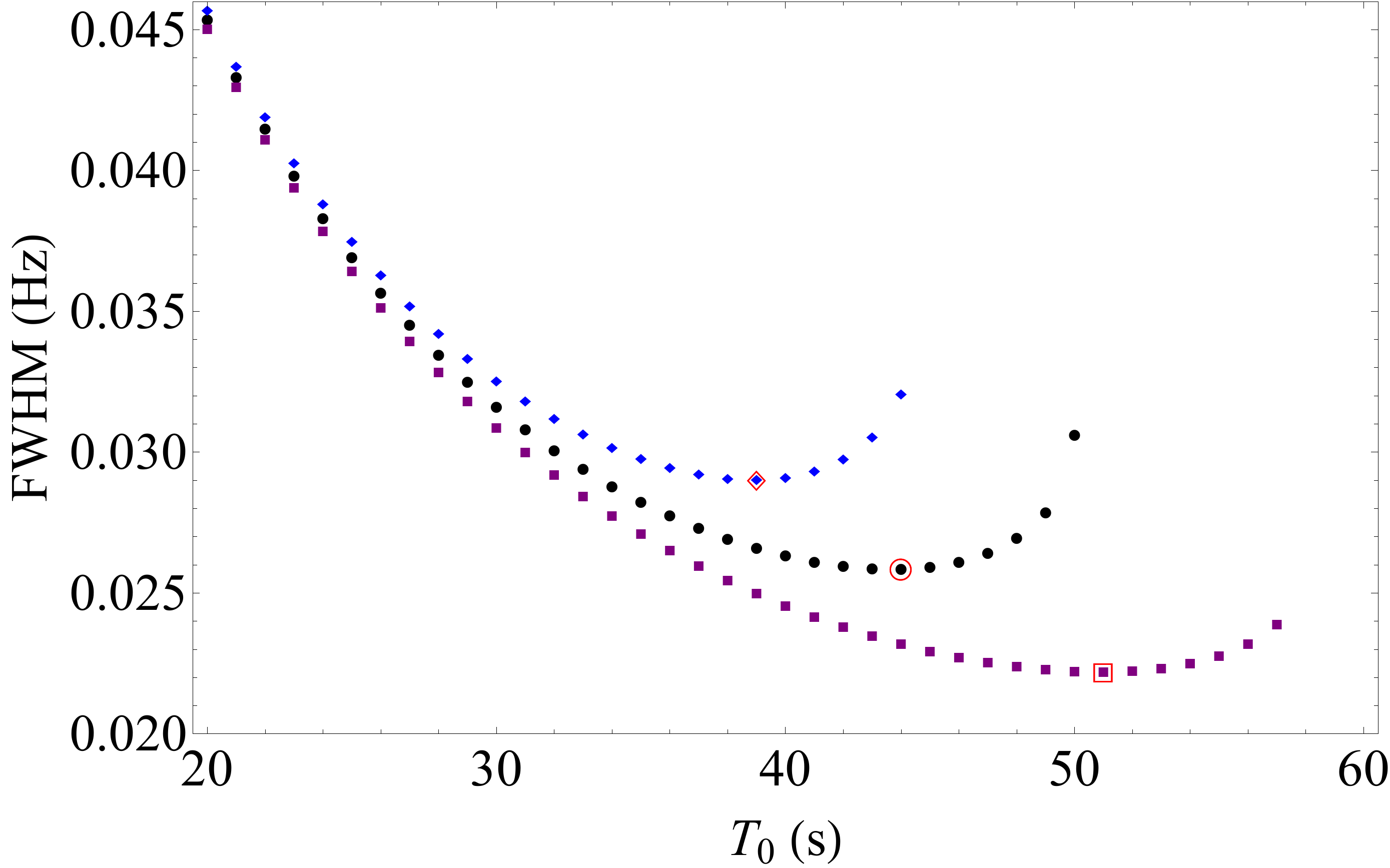}}
 \leavevmode
 \caption{ (Color online)   The numerically calculated FWHM of the $\alpha$-laser as a function of observation time.  Black circles represent calculations using the model $S_\nu(f)$ that is quoted in the main text, excluding narrow resonance features that do not contribute to the FWHM.  Blue diamonds(purple squares) represent calculations using the upper(lower) limit quoted for $S_\nu(f)$.  The minimum value for each condition is outlined in red.  These values determine the minimum observable linewidth for the $\alpha$-laser: 26(4) mHz.   }   \label{fig1s}
\end{figure}

 \begin{figure}
{\includegraphics[width =1\linewidth]{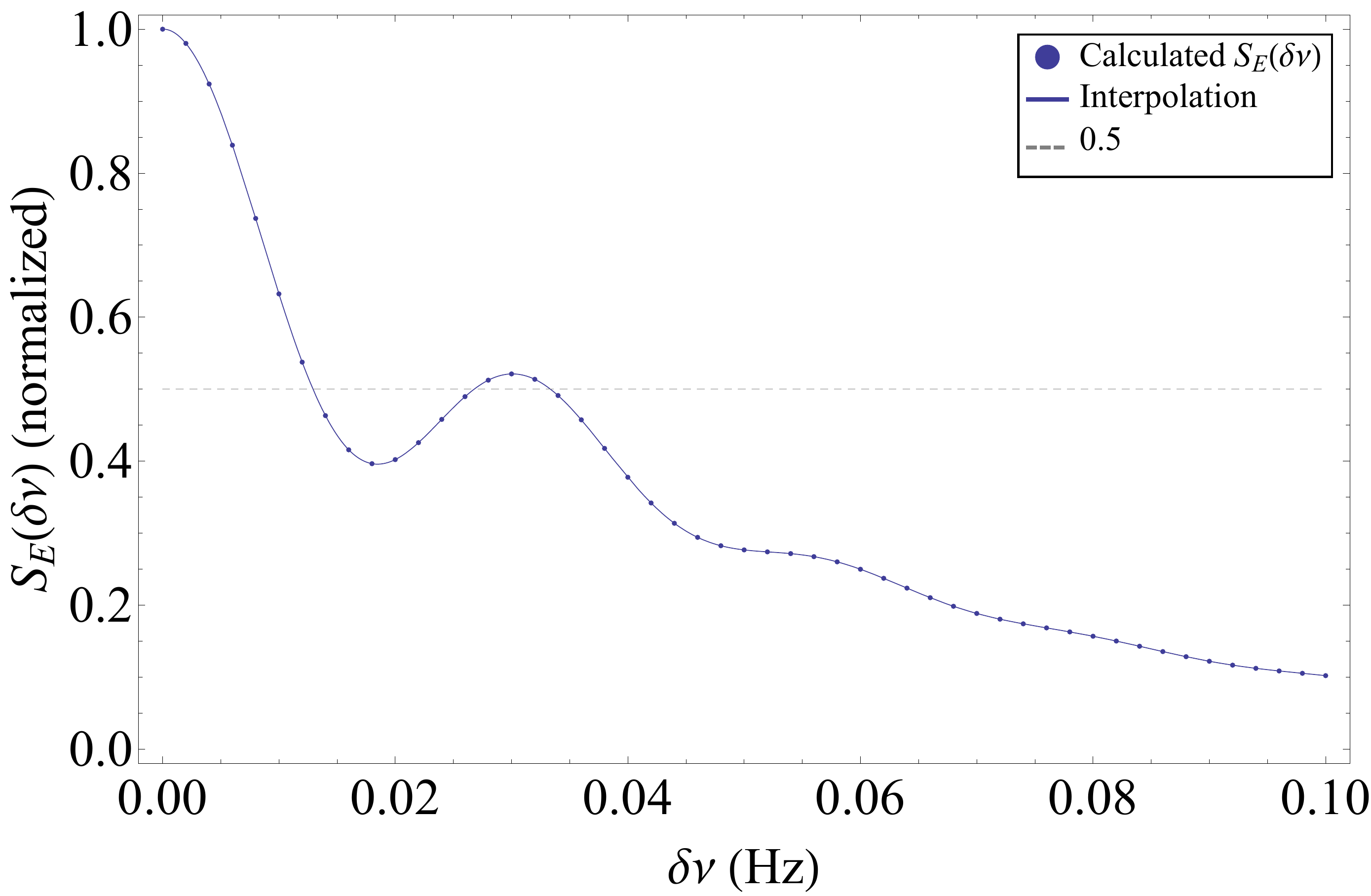}}
 \leavevmode
 \caption{ (Color online)   The calculated line shape, $S_E(\delta\nu)$, using h$_\mathrm{white}=0.0033$ Hz${}^2/$Hz, h$_\mathrm{thermal}=0.0015$ Hz${}^2$, and $T_0=44$ s.  This measurement time corresponds to the minimum FWHM of $S_E(\nu)$ for these values of $S_\nu(f)$.  A dashed gray line signifies the half maximum to guide the eye.}   \label{fig2s}
\end{figure}

To calculate the linewidth of the $\alpha$-laser, we express the autocorrelation function for the laser's electric field, $R_E(\tau)$, in terms of the frequency noise PSD, $S_\nu(f)$, and take the Fourier transform of $R_E(\tau)$ to get the laser line shape function \cite{elliott82,didomenico10}.  We consider an electric field of the form
\begin{equation}
E(t) = E_0e^{i(2\pi\nu_{L_0}t+\delta\phi(t))},
\label{eq:1c}
\end{equation}
where $\nu_{L_0} \equiv \omega_{L_0}/2\pi$
Then the $R_E(\tau)$ can be expressed in terms of $S_\nu(f)$ as follows: \cite{didomenico10}
\begin{equation}
R_E(\tau) = E_0^2e^{i2\pi\nu_{L_0}\tau}e^{-2\int_0^\infty S_\nu(f)\frac{\sin^2(\pi f \tau)}{f^2}df}.
\label{eq:2c}
\end{equation}
Since the model $S_\nu(f)$ that we deduce from our measurements contains a component proportional to $1/f$, the integral in the exponent diverges and we must take into account the measurement time over which the laser is observed.  A measurement lasting $T_0$ s is insensitive to frequency noise below $1/T_0$ Hz.  This dictates that the lower bound of the integral in Eqn.\ (\ref{eq:2c}) should be set to $1/T_0$, thereby making the integral finite.  The finite observation time also scales $R_E(\tau)$ by a triangle bump function since the measurement is windowed by a flat rectangle function between $t=0$ and $t=T_0$.  Therefore, the line shape observed over a finite time $T_0$ will depart from a simple Fourier transform of $R_E(\tau)$ and take the form
\begin{equation}
S_E(\nu) = 2\int_{-T_0}^{T_0}\left(1-\frac{\left|\tau\right|}{T_0}\right)e^{-i2\pi\nu\tau}R_E(\tau)d\tau,
\label{eq:3c}
\end{equation}
which can be simplified to
\begin{equation}
\begin{array}{rl}
S_E(\delta\nu) =& 4E_0^2\displaystyle\int_{0}^{T_0}\left(1-\frac{\left|\tau\right|}{T_0}\right)\cos(2\pi\,\delta\nu\,\tau) \\
\\
\times&e^{-2\int_{1/T_0}^\infty S_\nu(f)\frac{\sin^2(\pi f \tau)}{f^2}df}d\tau,
\end{array}
\label{eq:4c}
\end{equation}
where $\delta\nu \equiv \nu - \nu_{L_0}$.

To calculate the minimum observable FWHM, we compute the FWHM of $S_E(\delta\nu)$ as a function of $T_0$ for the model $S_\nu(f)$ quoted in the main text and its upper and lower bounds.  The three conditions lead to three minimum linewidths, observable at three different measurement times (Fig.\ \ref{fig1s}) and determine the minimum observable linewidth of the $\alpha$-laser to be 26(4) mHz.  The calculated line shape, $S_E(\delta\nu)$, using h$_\mathrm{white}=0.0033$ Hz${}^2/$Hz and h$_\mathrm{thermal}=0.0015$ Hz${}^2$ is plotted in Fig.\ \ref{fig2s} for the measurement time that give the minimum FWHM, $T_0=44$ s.  In practice we only include the white, 1/f, and broad acoustic resonance \#5 contributions to $S_\nu(f)$ for this calculation because the narrow peaks increase computation time and we have confirmed that these peaks do not significantly affect the calculated FWHM ($<0.1$\% effect for tested times).  For $T_0<30$ s,  the FWHM is dominated by the Fourier limit of the triangular windowing function (FWHM $=0.8859/T_0$ for $S_\nu(f) = 0$) with a small contribution from white noise ($h_\mathrm{white}=0.0033(3)$ Hz${}^2/$Hz leads to a Lorentzian linewidth of 10(1) mHz).  Between $T_0\cong40$ s and $T_0\cong60$ s, the line shape can cross 50\% of its peak value multiple times: a result of side bumps in the line shape arising from the windowing function and from the increasing contribution of 1/f noise as $T_0$ increases.  We always calculate the FWHM from the lowest frequency half maximum crossing.

\section{Technical comments}

In this work, we have used an optical heterodyne beat to verify the optical origin of noise observed in the excitation of an atomic system.  Any process that shifts the frequency of the atomic resonance will degrade the sensitivity of our measurement over the frequency range that it occurs, but distinguishing between laser and atomic noise processes is possible even without comparison to an optical beat.  For example, inhomogeneous atomic noise will be associated with reduced contrast in the spectroscopic signal.  Furthermore, noise that affects the atomic coherence will depend on the internal state of the atom.  For example, magnetic field noise will affect different nuclear spin states proportional to the projection of the nuclear spin along the quantization axis, whereas laser noise will affect all nuclear spin states equally.  Known processes that could lead to atomic noise, such as magnetic field noise, are limited to the $10^{-16}$ fractional frequency level at Fourier frequencies below a few hertz using the monitored central frequency and Zeeman splitting between nuclear spin levels during previous clock operation.

The fundamental limit to the accuracy of our spectrum analysis technique combines both the timing accuracy of the pulse sequences we use and our ability to reproducibly measure the amplitude of PSD for a given source of noise.  For the current work, the accuracy of our pulse sequences is limited by the reference oscillator of our experimental timing card which is at the 100 parts per million level.  When using longer total spectroscopy times, and thereby more precisely determining the frequency of measured noise peaks, we can use a more precise external reference to improve the accuracy of our pulse timing.  The accuracy of our measured PSD amplitude is determined by several factors: fluctuations in the actual noise present in our laser, uncertainty in the determination of the spectroscopic contrast, and the statistical uncertainty of our measurement.  Our measurements are generally repeatable at the level of $5\%$ over the course of a few days, after which both the optical beat measurements and atomic spectral analysis measurements show larger fluctuations in the amplitude of the discrete noise resonances. However, further work is needed to rigorously access the accuracy of our technique using a known noise source over a long period of time.

We note that stabilized frequency combs have demonstrated that the inherent noise associated with the mode locking process leads to relative noise between comb teeth at the $\mu$Hz level or below \cite{martin09}.  Furthermore, transfer of optical coherence between 1.5 $\mu$m and 698 nm has been demonstrated with $10^{-16}$ fractional frequency instability, limited only by the 1.5 $\mu$m laser \cite{hagemann13}.  Therefore, there is no fundamental limit to the sensitivity of our method of spectrum analysis arising from the transfer of phase coherence from lasers at other wavelengths to 698 nm using a frequency comb and our technique should be immediately applicable to sources at other wavelengths using standard phase lock techniques.

While we have proposed causes of the measured noise peaks, we do not have conclusive evidence as to their origin and it is possible that the 30 Hz and 22 Hz noise peaks are related due to their shared doublet structure.  For noise near 30 Hz we have proposed that alternating current motors, which exist in the building that houses our lab, are the cause.  Vibrations from these motors commonly put noise to our lasers even after several layers of heavily mechanically stabilized platforms.  Each alternating current motor will have a slightly different operating frequency due to the varying lag of different motors.  This could explain the doublet.    In reference to the 22 Hz noise peaks, we note that this frequency is near the acoustic resonance frequency of the room that houses the $\alpha$-laser, although this does not immediately explain the doublet structure.  We also note that since not all noise features exhibit this doublet structure, we believe that this structure does not arise from a flaw or limitation in our optical heterodyne measurement.  The strongest evidence lies in a direct and independent experimental measurement: the 22 Hz and 30 Hz doublets are present in vibration noise PSD measured on the table that supports the alpha-laser, which is actively vibration isolated, as well as in acoustic PSD measured inside the acoustic isolation box that houses our laser system.  The 24 Hz resonance observed in the optical heterodyne beat between the $\alpha$- and $\beta$-lasers has been previously observed in a beat between two nearly identical $\beta$-lasers.  It is believed to be a mechanical resonance of the support structure on which the reference cavity for the $\alpha$- laser is mounted.

\section{Table of parameters for model PSD features}

\begin{table}[h]
\caption{\label{table1}$S_\nu(f)$ Resonant Features.}
\begin{ruledtabular}
\begin{tabular}{c|c|c|c}
Index & $f_i$ (Hz) & $h_i$ (Hz${}^2/$Hz)& $\Gamma_i$ (Hz)\\ \hline
1 & 21.87 & 1.2 \cite{note2}  & 0.03 \\
2 & 22.39 & 0.6 \cite{note2}  & 0.03 \\
3 & 29.45 & 0.15 & 0.1  \\
4 & 29.90  & 0.08 & 0.4 \\
5 & 60 & 0.012 & 27 \\
\multicolumn{4}{c}{Added Noise} \\ \hline
6 & 46.6 & 0.07 & 2 \\
7 & 93.2 & 0.25 & 2 \\
\end{tabular}
\end{ruledtabular}
\end{table}